\documentclass[twocolumn,aps,pra,showpacs,
fixfloats]{revtex4}
\usepackage{bm}
\usepackage{dcolumn}
\usepackage{graphicx}
\begin{document}
\title{Revivification of confinement resonances in the photoionization of
$A$@C$_{60}$ endohedral atoms far above thresholds
}
\author{V. K. \surname{Dolmatov},
\email[To whom correspondence should be sent:]{vkdolmatov@una.edu}
G. T. \surname{Craven},
E.  \surname{Guler},
and
D. \surname{Keating}}

\affiliation{Department of Physics
and Earth Science, University of North Alabama, Florence, Alabama
35632, USA}

\date{\today}
\begin{abstract}
It is discovered theoretically that significant confinement resonances
in an $nl$ photoionization of a \textit{multielectron} atom $A$ encaged
in carbon fullerenes, A@C$_{60}$, may re-appear and be strong at photon
energies far exceeding the $nl$ ionization threshold, as a general
phenomenon. The reasons for this phenomenon are unraveled.
The Ne $2p$
photoionization of the endohedral anion Ne@C$_{60}^{5-}$ in the photon
energy region of about a thousand eV above the $2p$ threshold is chosen
as case study.
\end{abstract}
\maketitle

Endohedral fullerenes $A$@C$_{60}$, where an atom $A$ is confined
(encaged) inside a hollow carbon cage C$_{60}$, are modern frontline
targets of research in chemistry and physics. This is in view of their
novelty in basic science and importance to various applied sciences and
technologies. In particular, many efforts have been undertaken to
unravel trends in the response of $A$@C$_{60}$ confined atoms $A$,
referred to as $A$@ for brevity, to external perturbations, such as the
incoming photoionizing radiation (see Refs.
\cite{DolmAQC09,PranawaXe09,Himadri08,Amusia08,MuellerPRL08} and
references therein) and fast charged-particles \cite{GOS09}. One of
outstanding inherent features of the corresponding spectra is the
presence of resonances, termed confinement resonances
\cite{DolmAQC09,PranawaXe09,Amusia08}. New aspects of confinement
resonances are the subject of this paper.

Much of the current understanding of the nature and origin of
confinement resonances in spectra of confined atoms $A$@C$_{60}$ is
based on modeling the C$_{60}$ cage by a short-range attractive
spherical potential $V_{c}(r)$ of inner radius $r_{0}=5.8$ a.u., depth
$U_{0}=-8.2$ eV, and either the zero thickness, i.e.,
$V_{c}(r) = -U_{0}\delta(r-r_{0})$ \cite{Amusia08}, or finite thickness
$\Delta= 1.9$ a.u.
\cite{DolmAQC09,PranawaXe09,GOS09,MirrorCollapse09,LyrasJPB05} (and
references therein):
\begin{eqnarray}
 V_{c}(r)=\left\{\matrix {
-U_{0} <0, & \mbox{if $r_{0} \le r \le r_{0}+\Delta$} \nonumber \\
0 & \mbox{otherwise.} } \right.
\label{eqVc}
\end{eqnarray}
The formation of the $A$@C$_{60}$ system is completed by placing the
neutral atom $A$ at the center of the cage. For small sized, compact
atoms there is no charge transfer to the cage, so that the confined atom
$A@$ retains the structure of the free atom $A$. In the framework of
such modeling, confinement resonances in partial $nl$ ionization cross
sections of the $A$@ atom occur due to the interference of the ejected
photoelectron waves emerging directly from the confined atom, and those
scattered off the confining C$_{60}$ cage.

According to the thus accumulated database of calculated data,
confinement resonances in a partial $nl$ ionization cross section of a
$A$@C$_{n}$ system have been known to rapidly vanish with increasing
energy of the outgoing photoelectron, ceasing to exist at only some tens eV
above the $nl$ threshold, not to mention thousands eV above the
threshold. This is in line with a theory of scattering of particles off
a potential well/barrier. Indeed, starting from a sufficiently high
energy of the outgoing electron, the corresponding coefficient of
reflection off a finite potential well/barrier decreases with
increasing energy of the electron. As a result, the interference effect
between the outgoing and scattered electron waves becomes weaker, with
increasing energy of the electron, and so are the associated confinement
resonances. For the case of $A$@C$_{60}$ photoionization, the confining
potential, Eq.\ (\ref{eqVc}), and, thus, the C$_{60}$ cage itself,
become invisible to an outgoing $nl$ photoelectron at energies that are
only some tens eV above the $nl$ threshold. This is because the confining
potential $V_{c}(r)$ is shallow, being only a few eV deep. Consequently, the $nl$
photoionization cross sections of the confined and free atoms become
virtually identical at these energies, and they previously have been
thought to remain nearly identical at all higher energies.

We show in this paper that, contrary to the existing
understanding of the behavior of confinement resonances as a function of
ejected photoelectron energy, the importance of the shallow confining
potential, or the cage itself, generally reemerges at high photoelectron
energies, for confined \textit{multielectron} atoms. As a result,
confinement resonances manifest, i.e., revive, in the $nl$
photoionization spectra of a confined multielectron atom $A@$ in a region far
above the threshold energy. We term this effect the
\textit{revivification of confinement resonances}. This revivification causes
the corresponding
spectra of the free and confined atom be much different from each other,
as they are near threshold.  The prediction
and study of the revivification of confinement resonances effect constitutes
the quintessence of the present
paper. To qualitatively explain, and quantitatively study this effect,
we consider the $2p$
photoionization of Ne@ from a quintuply-charged endohedral fullerene
anion Ne@C$_{60}^{5-}$, near the Ne@ $1s$ threshold; we deem this case
study a most illustrative one among other possibilities.

To calculate the one-electron set of electronic bound and continuous
wavefunctions and energies, as well as photoionization amplitudes of
thus confined Ne, we follow, step-to-a step, the methodology
described in details in \cite{DolmNe@C60z}. In short, the C$_{60}$ is
simulated by the potential $V_{c}(r)$, Eq.\ (\ref{eqVc}). The excessive
negative charge $q=-5$ on C$_{60}$ is evenly distributed over the entire
outer spherical surface of the cage. The excessive charge on the
C$_{60}$ cage brings up an extra Coulomb potential $V_{q}(r)$ in
addition to the neutral cage potential $V_{c}(r)$:
\begin{eqnarray}
V_{q}(r)=\left\{\matrix {
\frac{q}{r_{0}+\Delta}, & \mbox{if $0 \le r \le r_{0}+\Delta$} \nonumber \\
\frac{q}{r} & \mbox{otherwise.} } \right.
\label{eqVq}
\end{eqnarray}
The sum total of these two potentials, $V(r) = V_{c}(r) + V_{q}(r)$, is
added to nonrelativistic Hartree-Fock (HF) equations for a free closed
shell atom, thereby turning the ``free atom'' HF equations
\cite{Amusia-book} into ``confined'' HF equations. Calculated HF
electronic energies and wavefunctions of the confined Ne atom are used
for calculating dipole $nl$ photoionization amplitudes of the atom. To
account for interchannel interaction/coupling in the Ne@
photoionization, the \textit{random phase approximation with exchange}
(RPAE) \cite{Amusia-book} is utilized with HF employed
as the zero-order approximation. RPAE has proven to be
a very reliable methodology over the years \cite{Amusia-book}.
With the thus calculated RPAE photoionization amplitudes
and their phase shifts, dipole angle-integrated partial photoionization
cross sections $\sigma_{nl}^{Ne@}(\omega)$ and dipole angular
asymmetry-parameters $\beta_{nl}^{Ne@}(\omega)$ of photoelectrons
ejected from Ne@ are determined. The corresponding expressions
for these quantities of an $A@$ photoionization are exactly the same as
those for any free closed shell atom $A$, see, e.g., Ref.\
\cite{Amusia-book}. When accounting for interchannel
coupling in the Ne@ $2p$ photoionization, RPAE calculations have to
include coupling between all possible photoionization channels
$1s$ $\rightarrow$ $p$,
$2s$ $\rightarrow$ $p$,
$2p$ $\rightarrow$ $d$,
and $2p$ $\rightarrow$ $s$,
since it is expected, on the basis of free Ne data \cite{Pranawa-Ne},
that none of these channels can be discarded in the photon energy region of
interest. Finally, in RPAE calculations, we use HF calculated
ionization thresholds instead of experimental ones, just for the sake of
``theoretical'' consistency. For $2p$, $2s$, and $1s$ ionization
thresholds of free Ne these are $I_{1s}^{Ne} = 894$, $I_{2s}^{Ne} = 53$,
and $I_{2p}^{Ne} = 23$ eV versus $I_{1s}^{Ne@} = 874$, $I_{2s}^{Ne@} =35
$, and $I_{2p}^{Ne@} = 5.5$ eV for Ne confined in C$_{60}^{5-}$. A
notable difference between the $I_{nl}$ and $I_{nl}^{Ne@}$ ionization
thresholds is due chiefly to the Coulomb potential brought about inside
the C$_{60}^{5-}$ cage by the excessive negative charge on the cage,
Eq.\ (\ref{eqVq}).

\begin{figure}[htbp]
\center
\includegraphics[width=8cm] {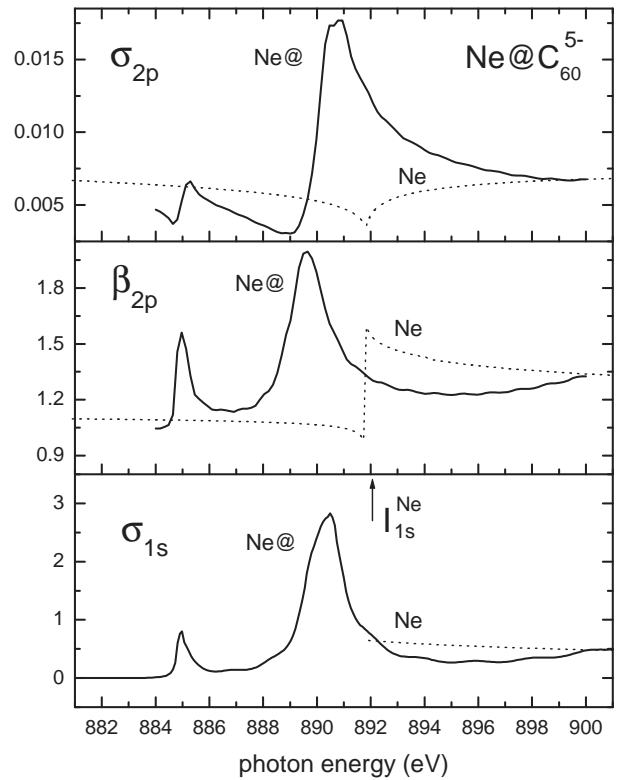}
\caption{
RPAE data for the partial angle-integrated $2p$ photoionization
cross section (Mb) (upper panel) and dipole $2p$ photoelectron
angular-asymmetry parameter (middle panel) of confined Ne@C$_{60}^{5-}$
(solid lines) and free Ne (dotted lines). Lower panel: RPAE
data for the $1s$ photoionization cross section $\sigma_{1s}^{Ne@}$ (Mb)
of confined Ne@C$_{60}^{5-}$ (solid line) and free Ne (dotted line).
Note that HF calculated ionization thresholds (see the main text) were used
in the calculations. If experimental thresholds were used, the calculated
curves would have been shifted by about $22$ eV towards lower energies.
This is because the HF ($I_{1s}^{Ne} \approx 892$ eV) and experimental
($I_{1s}^{Ne} \approx 870$ eV) values of the $1s$ ionization threshold
of free Ne differ from each other by about $22$ eV, and so the difference is
expected to be in similar Ne@.} \label{Fig1}
\end{figure}

RPAE results for $\sigma_{2p}^{Ne@}(\omega)$ and
$\beta_{2p}^{Ne@}$, above the $1s$ threshold, are depicted in Fig.\
\ref{Fig1} along with corresponding data for free Ne.
One can see that both $\sigma_{2p}^{Ne@}$ and $\beta_{2p}^{Ne@}$ show
the two noticeable resonances emerging above the $I_{1s}^{Ne@}$
threshold ($\approx 874$ eV), whereas there is nothing even remotely
similar in $\sigma_{2p}^{Ne}$ and $\beta_{2p}^{Ne}$ of free Ne. Thus,
the confinement, i.e., the charged C$_{60}$ cage itself, represented by
the sum total of the potentials $V_{c}(r)$ and $V_{q}(r)$,
surprisingly starts mattering again, and quite appreciably, at the
$2p$ photoelectron energy which is of the order of a thousand eV above
the small confining potential; this, as well, may be termed as
\textit{reemerging confinement effect}.
 Hence, the resonances in
$\sigma_{2p}^{Ne@}$ and $\beta_{2p}^{Ne@}$ are confinement resonances in
origin, since they are brought about by the confinement, thereby
illustrating the revivification of confinement resonances in the
high energy region of the $2p$ spectrum.

A hint to the physics behind the nature and origin of the revivification of
confinement resonances becomes evident when one explores the
earlier calculations \cite{DolmNe@C60z} of the $1s$ photoionization
cross section $\sigma_{1s}^{Ne@}$ of Ne@C$_{60}^{5-}$. The latter is
depicted in lower panel of Fig.\ \ref{Fig1} along with the corresponding
photoionization cross section of free Ne. The two resonances in
$\sigma_{1s}^{Ne@}$ are known to be confinement resonances
\cite{DolmNe@C60z}. Their presence in $\sigma_{1s}^{Ne@}$ is not
surprising, at these energies; they emerge only about $20$ eV
above threshold. There, the probability of reflection of $1s$
photoelectrons from the confining potential is appreciable, thereby
causing the emergence of confinement resonances in the $1s$
photoionization two of which, the strongest ones, are depicted in Fig.\
\ref{Fig1}. With this understanding of the $1s$ photoionization in mind,
the nature of the two resonances in the corresponding $2p$
photoionization of Ne@ can readily be unraveled. Specifically, when the photon
energy exceeds the $1s$ threshold energy, the $1s$ $\rightarrow$ $p$
photoionization channel opens, and it is dominated by
confinement resonances near threshold, as seen above. These confinement resonances are
further ``funneled'' to the $2p$ $\rightarrow$ $s$, $d$ channels, $via$
interchannel coupling. Given that the $\sigma_{2p}^{Ne@}$
photoionization cross section is much smaller than $\sigma_{1s}^{Ne@}$,
at these energies, the ``funneled'' $1s$ confinement resonances
show up strongly as confinement resonances in the Ne@ $2p$
photoionization as well. Hence, the significance of the confining cage
reemerges once again, via interchannel coupling with the $1s$ channel,
for high energy photoelectrons. As a consequence, confinement resonances
revive in the high energy region of the $2p$ photoelectron spectrum,
thereby driving it significantly away of that of the free atom. To test
this conclusion, we have performed a trial calculation with the $1s$
$\rightarrow$ $p$ channel excluded from RPAE calculations. The results
of the trial calculation (not shown) showed no sign
of any confinement resonances at all in the $2p$ photoionization in the
discussed energy region, in accordance with the above conclusion.

Note the fact that the $1s$ $\rightarrow$ $p$ channel couples strongly
with the $2p$ $\rightarrow$ $s$, $d$ channels near the $1s$ threshold is
not a new idea - for $2p$ photoionization of free Ne it
was illustrated earlier in \cite{Pranawa-Ne}. Nor is it new that confinement
resonances can show up in photoionization
cross sections of outer electrons through interchannel coupling. This
was demonstrated earlier for Xe $5s$ photoionization near the
Xe $4d$ threshold of the Xe@C$_{60}$ system \cite{PranawaXe09,DolmXe08},
where the confinement resonances, induced by interchannel coupling,
emerge along with the ``conventional'' Xe $5s$ confinement resonances
(i.e., confinement resonances which are not associated with interchannel coupling).
Nevertheless, in the Xe@ case, the correlation confinement resonances emerged
in a region of the $5s$ spectrum where conventional confinement resonances were
still expected.
The core novelty of the present paper versus Refs.\
\cite{PranawaXe09,DolmXe08} is that we find that confinement resonances
reemerge in $A$@C$_{60}$ valence shell
confined atom spectra about a thousand eV above threshold, far far above
where conventional wisdom said they would exist.

The discovered revivification of
confinement resonances effect appears to be an inherent feature of the
photoionization of confined \textit{multielectron} atoms exclusively,
whereas in confined \textit{single-electron} atoms, obviously, only the
``conventional'' confinement resonances may emerge. Furthermore,
the heavier the multielectron atom $A$, the greater the energy
difference between inner and outer subshells in the atom. Therefore,
when certain conditions are met (the presence of sizable confinement
resonances in inner shell channels along with their strong coupling with
outer shell channels), confinement resonances in outer shell spectra of
$A$@C$_{60}$ may revive even at tens keV above threshold. And, clearly, the
same type of confinement resonances may emerge in other types of
ionization processes, such as the ionization of $A$@C$_{60}^{z}$ atoms
by fast charged-particles, as a general phenomenon. Accordingly, a
revised understanding of the behavior of various types of ionization
spectra of an $A$@C$_{60}$ confined atom with increasing photon energy
should be adopted. Namely, initially, on the scale of tens of eV  above
threshold, the amplitudes of the confinement resonances in the $nl$ photoionization cross section of
the confined atom
will be diminished to
nearly a zero. However, at higher photon energies, i.e., hundreds or
thousands of eV above threshold, the corresponding cross section must generally
start exhibiting confinement resonances again. This will happen at
photon energies which correspond to opening of inner shell photoionization
channels whose intensities exceed by far the intensity of transitions from
the outer subshell of the confined atom and which are strongly coupled with the
inner-shell transitions.

This work was supported by NSF Grant No. PHY-$0652704$ and a UNA CAS Grant.
\end{document}